# Conformational Proofreading: The Impact of Conformational Changes on the Specificity of Molecular Recognition

Yonatan Savir[1] and Tsvi Tlusty[1,*]

[1]Department of Physics of Complex Systems, the Weizmann Institute of Science, Rehovot, Israel, 76100
[*]Corresponding author. E-mail: tsvi.tlusty@weizmann.ac.il

**Abstract**

To perform recognition, molecules must locate and specifically bind their targets within a noisy biochemical environment with many look-alikes. Molecular recognition processes, especially the induced-fit mechanism, are known to involve conformational changes. This arises a basic question: does molecular recognition gain any advantage by such conformational changes? By introducing a simple statistical-mechanics approach, we study the effect of conformation and flexibility on the quality of recognition processes. Our model relates specificity to the conformation of the participant molecules and thus suggests a possible answer: Optimal specificity is achieved when the ligand is slightly off target, that is a conformational mismatch between the ligand and its main target improves the selectivity of the process. This indicates that deformations upon binding serve as a conformational proofreading mechanism, which may be selected for via evolution.

**Introduction**

Practically all biological systems rely on the ability of bio-molecules to specifically recognize each other. Examples are antibodies targeting antigens, regulatory proteins binding DNA and enzymes catalyzing their substrates. These and other molecular recognizers must locate and preferentially interact with their specific targets among a vast variety of molecules that are often structurally similar. This task is further complicated by the inherent noise in the biochemical environment, whose magnitude is comparable with that of the non-covalent binding interactions [1-3].

It was realized early that recognizing molecules should be complementary in shape, akin of matching lock and key (figure 1A). Later, however, it was found that the native forms of many recognizers do not match exactly the shape of their targets. There is a growing body of evidence for conformational changes upon binding between the native and the bound states of many biomolecules, for example in enzyme-substrate [4], antibody-antigen [5-9] and other protein-protein complexes [10,11]. Binding of protein to DNA is also associated with conformational changes, which may affect the fidelity of DNA polymerase [12-15], and similar effects were observed in the binding of RNA by proteins [16-18]. The induced deformation typically involves displacements of binding sites in the range of tens of angstroms [5,10,12,16,19,20]. To account for these conformational changes upon binding, the induced fit scheme was suggested. In this scheme, the participating molecules deform to fit each other before they bind into a complex (figure 1B). Another model, the pre-equilibrium hypothesis, assumes that the target native state interconverts within an ensemble of conformations and the ligand selectively binds to one of them (figure 1C).

The abundance of conformational changes raises the question of whether they occur due to biochemical constraints or whether they are perhaps the outcome of an evolutionary optimization of recognition processes. In the present work, we discuss the latter possibility by evaluating the effects of conformation and flexibility on recognition. To estimate the quality of recognition we use the common measure of *specificity*, that is the ability to discriminate between competing targets. Whether conformational changes and especially the induced-fit mechanism can provide or enhance specificity has been a matter of debate [14,21-25]. Various detailed kinetic schemes have been suggested and their potential effects on specificity have been discussed – however without direct relation to concrete conformational mechanisms. Here we examine these underlying effects of flexibility and conformational changes that may govern the rate constants and thus determine specificity. Our approach tries to elucidate some of these basic effects by introducing a simple statistical-mechanics model and applying it to a generalized kinetic scheme of recognition in the presence of noise. As an outcome, the flexibility of the ligand and its relative mismatch with respect to the target which optimize specificity can be evaluated.

In the binding schemes described above (figure 1), the ligand is a "switch" that interconverts between a native, inactive form and an active form that fits the target. However, in a noisy biochemical environment, one may expect both the ligand and the targets to interconvert within an ensemble of many possible conformations. Such an ensemble may be the outcome, for example, of thermally induced distortions. Consider for example a scenario in which an elastic ligand is interacting with two rigid competing targets (figure 2). All the conformations of the ligand may interact with the targets and as a result a variety of complexes, differing by the structures of the bound ligand, is formed (figure 2). Among the complexes formed, some are composed of perfectly matched ligand and target. In those complexes, specific binding energy due to the alignment of binding sites is gained. However, a complex may be formed even if the ligand does not perfectly match the target, due to non-specific binding energy. For example, the *lac* repressor can bind non-specifically to DNA regardless of its sequence [26]. All the complexes, the matched and the mismatched, may retain some functionality. The efficiency of the recognition process depends on the elasticity of the ligand and on the structural mismatch between the ligand native state and the main target.

The quality of recognition is measured by its specificity, which is defined as follows. Consider a ligand $a$ interacting with a correct target $A$ and an incorrect competitor $B$,

$$a + A \underset{}{\overset{K_A}{\rightleftharpoons}} aA \xrightarrow{v_A} correct\ product \qquad (1)$$

$$a + B \underset{}{\overset{K_B}{\rightleftharpoons}} aB \xrightarrow{v_B} incorrect\ product \qquad (2)$$

where $K_A$ and $K_B$ are the dissociation constants, and $v_A$ and $v_B$ are the turnover numbers. Specificity is naturally defined as the ratio of the correct production rate, $R_A = [aA] \odot v_A$, and the incorrect production rate, $R_B = [aB] \odot v_B$, where [] denotes concentration. Typically, the chemical step is the rate-limiting one and the complex formation reaction is therefore in quasi-equilibrium, $[aA] = [a][A]/K_A$, $[aB] = [a][B]/K_B$. Thus the specificity $\xi$ takes the form:

$$\xi = \frac{R_A}{R_B} = \frac{(v_A/K_A)[A]}{(v_B/K_B)[B]}. \qquad (3)$$

If however the ligand and the targets interconvert within ensembles of many possible conformations (figures 2-3), the specificity includes potential contributions from all possible complexes,

$$\xi = \frac{\sum_{i,j}(v_{A,ij}/K_{A,ij})[a_i][A_j]}{\sum_{i,j}(v_{B,ij}/K_{B,ij})[a_i][B_j]}, \qquad (4)$$

where $i$ and $j$ denote the conformations of the ligand and the target, respectively. $K_{ij}$ is the dissociation constant of the complex formed from the $i$-th ligand conformation and $j$-th target conformation and $v_{ij}$ is the turnover number of this complex.

Equation 3 and its generalization, equation 4, reflect the dependence of the specificity on both the concentrations of complexes, determined by the dissociation constants $K$, and on their functionality, determined by the turnover numbers $v$. These parameters, $K$ and $v$, depend on the flexibility and structure of the participant molecules. Evaluating this dependence allows us to estimate the optimal flexibility and structure similarity between the ligand and the main target.

## Results and discussion

**Lowest elastic mode model**

In essence, molecular recognition is governed by the interplay between the interaction energy gained from the alignment of the binding sites and the elastic energy required to deform the molecules to align. Motivated by deformation spectra measurements [27], we treat this interplay within a simple model that takes into account only the lowest elastic mode. This is a vast simplification of the many degrees of freedom that are required to describe the details of a conformational change. However, as we suggest below, this simplified model still captures the essence of the energy tradeoff. Modeling proteins as elastic networks was previously applied to study large amplitude [28] and thermal fluctuations [29,30] of proteins, and to predict deformations and domain motion upon binding [27,31]. These models are fitted with typical spring constants of a few $k_B T/Å^2$.

We consider first an elastic ligand interacting with a rigid target. Later, we discuss the case of deformable targets. The binding domain of the ligand is regarded as an elastic string on which $N$ binding sites are equally spaced (figure 4). The elastic deformation energy is described by harmonic springs that connect adjacent binding sites. In the native state of the ligand, the length of the binding domain is $l_0$. The ligand interacts with a rigid target on which $N$ complementary binding sites are equally spaced along a binding domain of length $s$. Binding is specific, that is a binding site of the ligand can gain binding energy $\varepsilon$ only by fastening to its complementary binding site on the target. The ligand-target interactions are relatively short-range and therefore binding energy is gained only if the complementary binding sites are at the same position.

The presence of the target may induce a deformation of the ligand that, in order to gain binding energy, shifts the binding sites to new positions. However, such deformation of the ligand costs elastic energy. The conformation of the ligand is determined by $N$ degrees of freedom, the $N$ positions of the ligand binding sites. We assume for simplicity that all the springs that connect adjacent binding sites on the ligand have the same spring constant. We consider here only the deformation mode of lowest elastic energy, in which the binding domain of the ligand is stretched or shrunk uniformly. Thus, we reduce the number of degrees of freedom from $N$ to two, the length of the deformed binding domain $l$, and the position of its edge (figure 4).

To evaluate the effect of conformational changes and flexibility on specificity one needs to estimate the concentrations and reaction constants in (4). Since all the reactions besides the product formation are assumed to be in equilibrium, we can regard each conformation of the ligand, specified by its length $l_i$, as a separate *chemical species* $a_i$. Thus we may apply the law of mass-action to each of the binding reactions $a_i + A \leftrightarrow a_i A$, and obtain the equilibrium constant $K_{iA} = [a_i][A]/[a_iA] \sim Z_i Z_A / Z_{iA}$, where $Z_i$, $Z_A$ and $Z_{iA}$ are the single-particle partition functions of the $i$-th ligand conformation, target and complex, respectively. The equilibrium constant is (see Methods):

$$K_{iA} \sim Z_i Z_A / Z_{iA} \sim \left(\delta(l_i - s_A) e^{N\varepsilon} + e^f\right)^{-1},$$

(5)

where $f$ is the non-specific free energy. The binding energy $\varepsilon$ and the non-specific free energy $f$ are in units of $k_BT$. The concentration of free ligand of length $l_i$ is proportional to the Boltzmann exponent of the distortion energy, $[a_i] \sim [a]\cdot\exp(-k/2(l_i - l_0)^2)$, where the effective spring constant $k$ is in units of $k_BT/\text{length}^2$ and $[a]$ is the total concentration of the free ligand. Although some preferred conformations may be catalyzed much faster than the others, the interconversion is assumed to be fast enough to still maintain this equilibrium distribution.

With the knowledge of how the rate constants depend on the conformation and the flexibility of the ligand, we analyze below the specificity to suggest a simple answer to the question raised above: What are the optimal geometry and flexibility that yield maximal specificity? The quality of a recognition process depends on two main properties of the participant molecules, their chemical affinity and the conformational match between them. To discuss the conformational effect, we consider a main or "correct" target $A$ and an "incorrect" competitor $B$ that differ in structure; their binding domains are of different lengths, $s_A$ and $s_B$. Chemical affinity is taken into account by assuming that the competing target $B$ has only $N–m$ interacting binding sites while the main one has $N$. We test the specificity of a ligand specified by a native state length $l_0$ and a flexibility $k$. We define the *mismatch d* as the difference between the ligand's native state length and the correct target's length, $d = l_0 - s_A$. We first examine the competition between two rigid "noiseless" targets and then discuss the noisy case. The generalization to more than two competing targets is straightforward.

**Recognizing noiseless targets**

Consider a ligand interconverting within an ensemble of conformations, each one with a different binding domain length $l_i$. This ligand interacts with two competing rigid targets that differ by their length $\Delta = s_A - s_B$. The ratio of the production rates due to unmatched and matched complexes is denoted by $r = v_{um} / v_m$ where the production rates of the correct and incorrect products are assumed to be equal, $v_A = v_B$. For the sake of simplicity, we assume that the target is in excess with respect to the ligand, $[A] \sim [A_{total}]$ and $[B] \sim [B_{total}]$, and that the concentrations of the competing targets are equal, $[A] = [B]$. Substitution of the equilibrium constant (5) into (4) yields the specificity (see Methods)

$$\xi = \frac{R_A}{R_B} = \frac{\left(1+e^{N\varepsilon-f}\right)e^{-kd^2/2} + \alpha r}{\left(1+e^{(N-m)\varepsilon-f}\right)e^{-k(d+\Delta)^2/2} + \alpha r}, \tag{6}$$

where $f$ is the non-specific free energy. The dimensionless parameter $\alpha \sim 1/(k^{1/2}g)$ is the ratio between the typical length scales, $k^{-1/2}$ of the elasticity and $g$ of the binding potential. Thus, we obtained in (6) the specificity $\xi$ as a function of the structural and energetic parameters: the difference between the target and the competitor Δ, the mismatch between the ligand and the target $d = l_0 - s_A$, the effective spring constant $k$ and the specific and non-specific binding energies. Below we examine this dependence to find the optimal ligand, specified by its mismatch $d$ (or its native state length $l_0$).

The specificity (6) is simply the ratio of the formation rates of correct and incorrect products, $R_A$ and $R_B$, respectively (figure 5). The correct production rate $R_A$, as a function of the mismatch $d$, is the sum of a Gaussian centered on $d = 0$, which accounts for the specific binding, and a uniform non-specific contribution. $R_A$ is therefore maximal at a zero mismatch. The incorrect production rate $R_B$ has the same uniform non-specific contribution and its specific contribution is now a Gaussian centered around $d = \Delta$, where it exhibits its maximum. The crossover where the specific and non-specific contributions become comparable defines a *"window of recognition"*. When the windows of recognition of the correct and incorrect targets overlap, the resulting specificity exhibits a maximum at a *finite nonzero mismatch* (figure 5A). This optimal mismatch $d_0$ is approximately

$$d_0 \simeq k^{-1/2}\left((N-m)\varepsilon - f - \log(\alpha r)\right)^{1/2} - \Delta. \tag{7}$$

As the ligand becomes more rigid, the specificity increases while the optimal mismatch $d_0$ tends to zero (figure 6A). The optimal mismatch is bounded by $d_{max} = (N\varepsilon/k)^{1/2}$, the length-scale that reflects the interplay between the elastic and specific binding energies.

Competing targets of similar structure $\Delta \approx 0$, have both correct and incorrect recognition windows centered on zero mismatch and the resulting specificity is akin to a rectangular window (figure 5B). The width of this window is the mismatch where specificity is half of its maximum, $d_{1/2} \approx k^{-1/2}((N-m)\varepsilon - f - \log(\alpha r))$. As the ligand becomes more flexible the width of this rectangular window increases (figure 6B). Targets that differ much are evidently not competitors. Indeed, if the difference $\Delta$ is much larger than the window of recognition, the optimal mismatch vanishes (figure 5C). Thus, (7) provides a criterion for relevant competitors: these must lie within the window of recognition of the correct target.

An interesting special case is when only the perfectly matched complexes are functional. This situation may occur if the non-specific binding energy is small and only matched complexes are formed, or if mismatched complexes are not functional, $r=0$. The specificity in this case increases exponentially with the mismatch, $\xi \sim \exp(k \cdot \Delta \cdot d)$ (figure 6C). Generalization of these results to more then 1D and for multiple competitors is straight forward. In figure 7, the specificity of a ligand as a function of mismatch in the presence of a few competitors is shown. The optimal mismatch depends on the structure of the various competitors. When the competitors have a structure similar to that of the main target, the mismatch is non-zero.

These results are reminiscent of kinetic proofreading [32,33] in which the specificity of a biochemical reaction increases exponentially with the *temporal delay* or the number of additional intermediate states [34-37]. In kinetic proofreading, the delay reduces the production rates of both correct and incorrect products, but the reduction of the incorrect product is larger and thus specificity improves. In the present case, the equivalent of the temporal delay is the *spatial mismatch*. It is evident from equation (6) that mismatch reduces both the correct and incorrect rates, but as the effect on the incorrect rate is more significant the overall specificity increases. Of course, a major difference is that kinetic proofreading is an energy-consuming non-equilibrium scheme whereas the *conformational proofreading* suggested here is at quasi-equilibrium.

**Recognizing noisy targets**

In this case, the ligand still interconverts between an ensemble of conformations, but now the target is prone to error. We describe this noise as Gaussian fluctuations of the target's length $s$ with a variance $\sigma$. These fluctuations may originate from various sources such as thermal noise, where the variance $\sigma$ is related to the target's flexibility as $\sigma_{A,B} \sim k_{A,B}^{-1/2}$. The noise introduces additional matched complexes and thus widens the windows of recognition of both the correct and incorrect targets. Similar to (6) (see Methods), the resulting specificity is

$$\xi = \frac{\left(1+e^{N\varepsilon-f}\right)e^{-kd^2/2(1+k\sigma_A^2)} + \alpha r(1+k\sigma_B^2)^{-1/2}}{\left(1+e^{(N-m)\varepsilon-f}\right)e^{-k(d+\Delta)^2/2(1+k\sigma_B^2)} + \alpha r(1+k\sigma_A^2)^{-1/2}}, \tag{8}$$

If $\sigma_A = \sigma_B = \sigma$, the results are the same as for two rigid targets competing for a ligand with an effective spring constant $k' = k/(1+k\sigma^2)$. For any values of $\sigma_A$ and $\sigma_B$, when the targets differ in structure $\Delta \neq 0$, the specificity is optimal at a nonzero mismatch as in the noiseless case (figure 8A-B). But unlike the noiseless scenario, even the specificity of an infinitely rigid ligand may be optimal at a nonzero mismatch. For $\Delta = 0$ the specificity has an extremum at a zero mismatch. If the incorrect target is noisier, $\sigma_A < \sigma_B$, identical ligand and target achieve maximal specificity (figure 8C-D). However, if the correct target is noisier $\sigma_A > \sigma_B$, a mismatched ligand is optimal *even for structurally similar targets*.

**Possible experimental tests**

The conformational proofreading model makes several predictions that may be put to an experimental test. To begin with, the structure of the target, the ligand and the competing molecule should fulfill a number of relations. First, we expect a mismatch to occur only if a competitor is within the ligand's window of recognition, since this is the situation where competition may threaten the quality of recognition. This can be verified by comparing the structure of the native ligand, its main target and the competitor. Second, we predict that a compromise must be struck between the need for the native ligand to be as far as possible from the competitor and as close as possible to the target, so that the mismatch will place ligand and competitor at opposite sides of some structural axis. For example, in figure 7 the mismatch which maximizes specificity is determined by the location of the competitors recognition windows. Experimentally, we expect that there will be a need for resolved 3D structures of ligands both in their native state and bound to their target, as well as these of the competitors. The rapidly increasing structural information that is available from studies of molecular recognition systems suggests that data that can validate or falsify our conformational proofreading hypothesis may already be available, or readily obtained.

Besides observing competition and specificity in known biological system, an experiment that in principle allows control over the nature of competition and the functional results of this competition may be carried out. A particularly appealing system that can be experimentally accessed and manipulated is that of transcription factors. While a transcription factor has one or several specific binding sites, there may be many competing sites on the DNA that would bind it. One can therefore experimentally alter the specific binding site or its competing sites, as well as the transcription factor, by point mutations and then observe the effect on specificity, e.g. by measuring the expression of upstream genes. The next step in this direction would be, instead of artificially manipulating the structures, tracing the coevolution of the transcription factor and all of the binding sites, looking at the *in-vivo* evolutionary optimization of recognition.

**Conclusion**

The ability to perform efficient information processing in the presence of noise is crucial for almost any biological system. Enhancing the specificity of recognition, in the sense of discrimination between competing targets, is therefore expected to increase the fitness. By introducing a model that captures the essence of the tradeoff between the specific binding energy and the structural deformation energy, it appears possible to estimate the optimal flexibility and geometry of the fittest molecules. Our model suggests that to optimally discriminate between competing targets of different structures, the ligand should have a *finite mismatch* relative to the main target. This spatial mismatch is similar to the temporal delay that underlies kinetic proofreading. Our analysis suggests that conformational changes upon binding may arise as the outcome of an evolutionary selection for enhancing recognition specificity in a noisy environment. This may also suggest that the structure and flexibility of binding molecules are governed by evolutionary pressure to optimize not only specificity but other cost functions such as robustness to noise.

## Methods

**Dissociation constant calculation**

Within the lowest mode model assumptions, when the ligand and the target are perfectly aligned, $x_0 = y_0$ and $l = s$, all the binding sites interact and contribute a total binding energy $N\varepsilon$. Otherwise, the binding energy is only due to a single interacting site. If there are many binding sites, we can neglect the single site contribution and approximate the interaction energy by:

$$H = \frac{1}{2}k(l-l_0)^2 - N\varepsilon \cdot \delta(x_0 - y_0) \cdot \delta(l-s), \qquad (9)$$

where $k$ is the effective spring constant of the ligand binding domain. The interaction energy (9) describes an idealized scenario in which only perfectly aligned ligand and target gain specific binding energy. Of course, in reality there could be other conformations with partial alignment, but they would require the excitation of higher elastic modes.

In order to calculate the partition function of the complex $Z_{iA} = \text{Tr}(\exp[-H(l_i)])$ all the possible binding configurations of this complex should be specified. As mentioned above, only in the perfectly aligned configuration the specific binding energy $N\varepsilon$ is gained. However, there may be other configurations in which the ligand and the target are bound non-specifically. We roughly estimate the non-specific contribution to the partition function as the product of the volume in which the non-specific binding occurs and the exponent of the non-specific binding energy. This non-specific contribution is $\exp(f)$, where $f$ is defined to be the non-specific free energy. The total complex partition function is the product of the elastic contribution and the contribution due to binding, specific and non-specific, $Z_{iA} = \exp(-k/2(l_i - l_0)^2) \cdot (\delta(l_i - s) \cdot \exp(N\varepsilon) + \exp f)$. The elastic contributions to (5) cancel out since they are equal for both the ligand and ligand-target partition functions. The irrelevant kinetic contributions were also omitted.

**Specificity of a ligand with continuous ensemble of conformations**

The ligand may interconvert within a continuous ensemble of conformations specified by their binding site length $l$. Since the complex formation reaction is in quasi-equilibrium, the concentration of free ligand of length $l$ is proportional to the Boltzmann exponent of the distortion energy, $[a(l)] \sim [a] \cdot \exp(-k/2(l - l_0)^2)$, where $l_0$ is the native state length. The effective spring constant $k$ is in units of $k_BT/\text{length}^2$ and $[a]$ is the total concentration of the free ligand. The dissociation constant in its continuous form is

$$K_{A,B}(l) \sim \left(\delta(l - s_{A,B})e^{N\varepsilon} + e^f\right)^{-1}. \qquad (10)$$

Only matched complexes in which $l = s_{A,B}$ gain specific binding energy and the turnover of these complexes, $v_m$, may be different from the turnover number of the unmatched complexes, $v_{um}$. Therefore, the continuous form of the turnover number is

$$v_{A,B}(l) = v_{A,B,m}\delta(l - s_{A,B}) + v_{A,B,um}(1 - \delta(l - s_{A,B})) \tag{11}$$

If the competing targets are rigid, the contribution to specificity from all possible complexes (4) becomes an integral over all ligand conformations $l$,

$$\xi = \frac{[A]\int_0^\infty (v_A(l)/K_A(l))[a(l)] \cdot dl}{[B]\int_0^\infty (v_B(l)/K_B(l))[a(l)] \cdot dl} \tag{12}$$

$$= \frac{\int_0^\infty e^{-k/2(l-l_0)^2} \cdot (\delta(l-s_A)e^{N\varepsilon} + e^f) \cdot (v_{A,m}\delta(l-s_A) + v_{A,um}(1-\delta(l-s_A)) \cdot dl}{\int_0^\infty e^{-k/2(l-l_0)^2} \cdot (\delta(l-s_B)e^{(N-m)\varepsilon} + e^f) \cdot (v_{B,m}\delta(l-s_B) + v_{B,um}(1-\delta(l-s_B)) \cdot dl}.$$

For the sake of simplicity we assume that (i) $v_{A,m} = v_{B,m}$ and $v_{A,um} = v_{B,um}$, (ii) the target is in excess with respect to the ligand, $[A] \sim [A_{total}]$ and $[B] \sim [B_{total}]$, and (iii) the concentrations of the competing targets are equal, $[A] = [B]$. Performing the integration yields

$$\xi = \frac{gv_m(e^{N\varepsilon} + e^f)e^{-\frac{k}{2}d^2} + e^f v_{um}\sqrt{\frac{\pi}{2k}}erf(1 + \sqrt{\frac{k}{2}}l_0) - gv_{um}e^f e^{-\frac{k}{2}d^2}}{gv_m(e^{(N-m)\varepsilon} + e^f)e^{-\frac{k}{2}(d+\Delta)^2} + e^f v_{um}\sqrt{\frac{\pi}{2k}}erf(1 + \sqrt{\frac{k}{2}}l_0) - gv_{um}e^f e^{-\frac{k}{2}(d+\Delta)^2}}, \tag{13}$$

where $d = l_0 - s_A$ and $\Delta = s_A - s_B$. The normalization factor $g$ reflects the assumption of a continuous ensemble of ligand conformations $l$. $g$ is the phase space cell volume (actually, the translational factor of this cell volume). This cell volume appears as a proportionality constant of the partition functions. $g$ is proportional to the typical length scale of thermal fluctuations in the system [38] which is affected by the elastic and binding forces. The $k$-dependence of $\alpha = (kg^2/2\pi)^{-1/2}$ is at most $\alpha \sim k^{-1/2}$ and therefore contributes only logarithmic correction in equations (6-8). Under the reasonable assumptions that $v_m >> v_{um}$ and $k^{1/2}l_0 >> 1$ the specificity becomes (6). The above assumptions are made for simplicity and clarity, they do not change the qualitative nature of the results.

If the targets are subject to noise in their structure, (12) should also be integrated over all possible target conformations. If the fluctuations of the target binding site are around native state lengths $s_A$ and $s_B$ with variances $\sigma_A$ and $\sigma_B$, the specificity is

$$\xi = \frac{\int\limits_0^\infty\int\limits_0^\infty e^{-k/2(l-l_0)^2} e^{-(s_A'-s_A)^2/(2\sigma_A^2)} \cdot (\delta(l-s_A)e^{N\varepsilon} + e^f) \cdot (v_m \delta(l-s_A) + v_{um}(1-\delta(l-s_A))) \cdot dl ds_A'}{\int\limits_0^\infty\int\limits_0^\infty e^{-k/2(l-l_0)^2} e^{-(s_B'-s_B)^2/(2\sigma_B^2)} \cdot (\delta(l-s_B)e^{(N-m)\varepsilon} + e^f) \cdot (v_m \delta(l-s_B) + v_{um}(1-\delta(l-s_B))) \cdot dl ds_B'}. \quad (14)$$

If again, we assume that $v_m \gg v_{um}$, $k^{1/2}l_0 \gg 1$ and $s_{A,B}/\sigma_{A,B} \gg 1$, performing the integral (14) yields (8).

**Figure Legends**

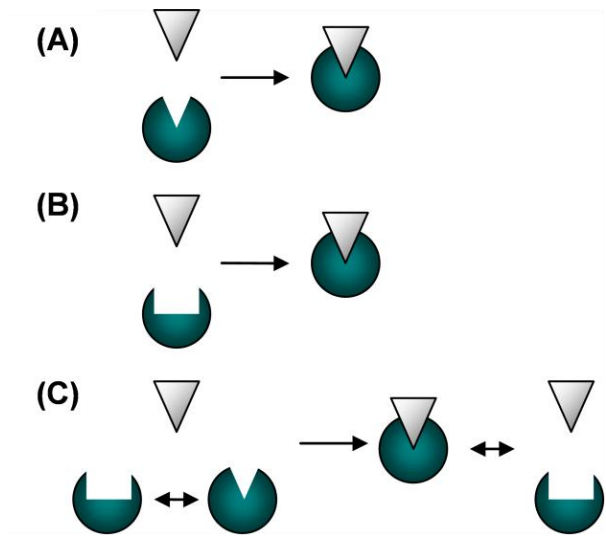

**Figure 1**: **Models of molecular recognition**. **(A)** Lock and key. No conformational changes occur upon binding. The ligand (white) and the target (green) have complementary structures. **(B)** Induced fit. The target changes its conformation due to the interaction with the ligand. **(C)** Pre-existing equilibrium model. The native state is actually an ensemble of conformations, that is deformations may occur even before binding. The ligand selectively binds the matching target within this ensemble of fluctuating conformations.

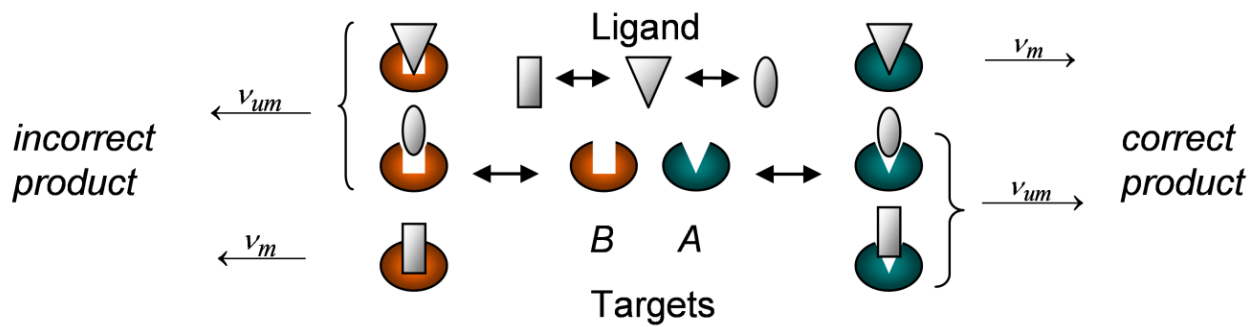

**Figure 2:** **Competition between two rigid targets**. The ligand (white) is interconverting within an ensemble of conformations while interacting with two rigid competitors, *A* and *B* (green and orange), characterized by different structures. Non-specific binding energy may lead to the formation of functional complexes in which the target and the ligand are not exactly matched. The unmatched complexes may also be functional but their product formation rates, $v_{um}$, may differ from these of the matched complexes, $v_m$. The specificity of the ligand, that is its ability to discriminate between *A* and *B*, depends on the ligand flexibility, the structural mismatch between its native state and the correct target and on the structural difference between the competing targets.

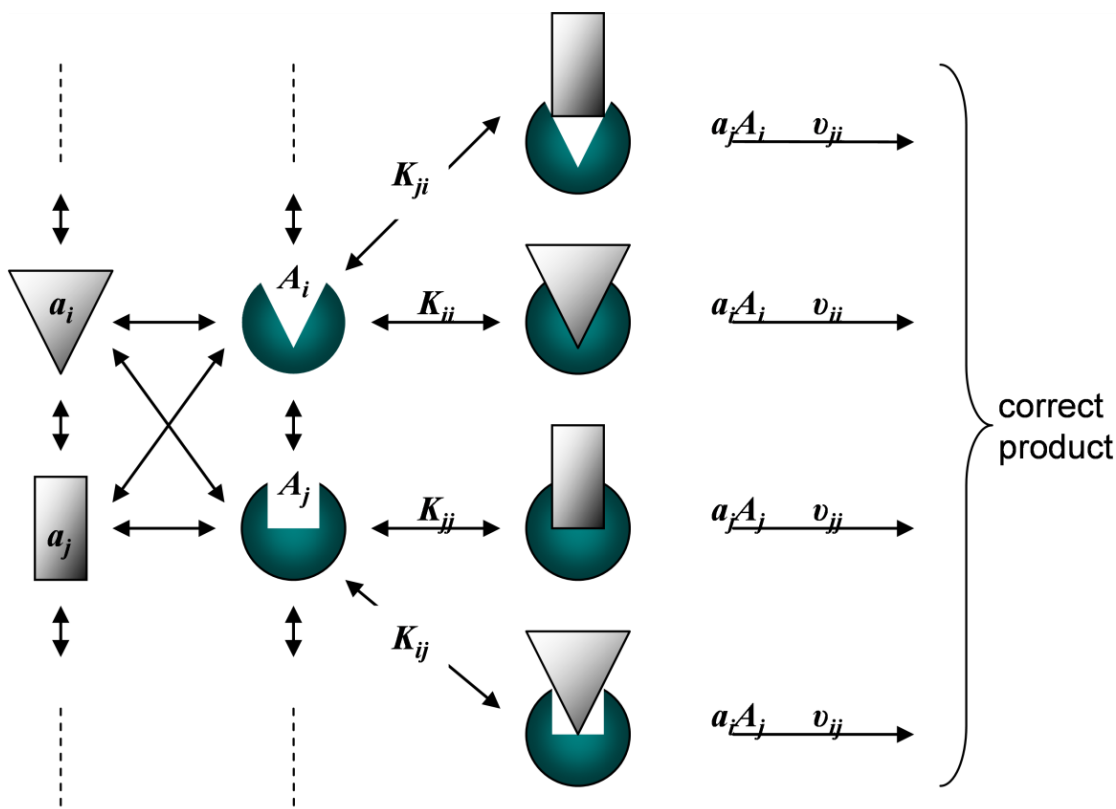

**Figure 3**: **General molecular recognition scheme.** Both the ligand (white) and the target (green) are interconverting between an ensemble of conformations denoted by indices, $a_i$ and $A_i$, respectively. All the different conformations may interact and as a result, a variety of complexes is formed. In some of them the target and the ligand are perfectly matched, for example $a_iA_i$ and $a_jA_j$, and in some there is only partial fit, for example $a_iA_j$ and $a_jA_i$. The rate of product formation depends on the concentrations of the complexes, which depend on $K_{ij}$, and on the functionality of each complex, which depends on the turnover numbers, $v_{ij}$. In a similar fashion, the different ligand conformations, $a_i$, may interact with competing target conformations $B_i$ and thus catalyze incorrect product.

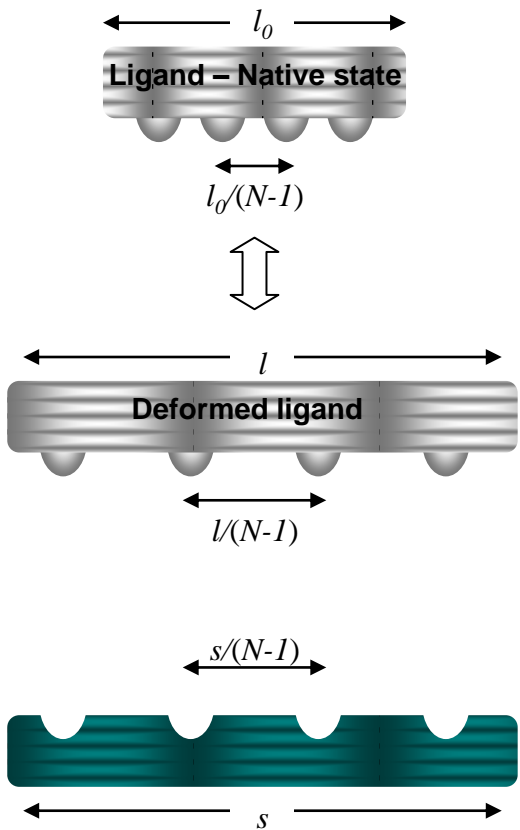

**Figure 4: Lowest elastic mode model.** Conformational changes occur upon binding of a ligand (white) and a target (green). In the native state of the ligand, the binding sites are equally spaced and positioned at $x_i^0$ ($i = 0,1,2…N-1$) and the total length of the binding domain is $l_0$. The ligand is interacting with a rigid target on which $N$ complementary binding sites are equally spaced and positioned at $y_i = y_0 + i \cdot s/(N-1)$ where $s$ is the length of the target binding domain. The ligand may undergo a conformational change to fit the target. Since we consider only the lowest mode motion, the ligand may only stretch or expand uniformly. Thus, its binding sites are displaced to $x_i = x_0 + i \cdot l/(N-1)$ and the total length changes from $l_0$ to $l$.

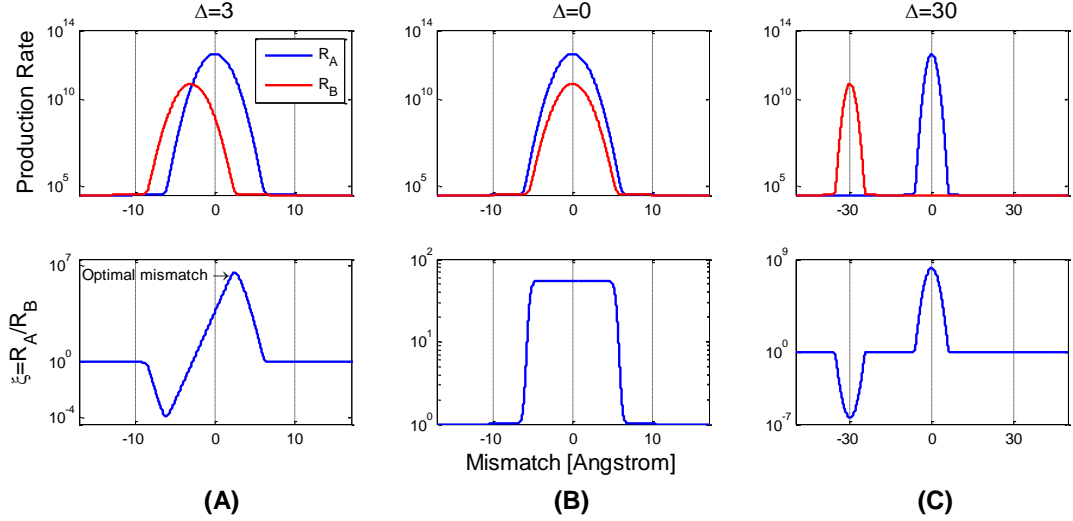

**Figure 5: The dependence of specificity on mismatch**. Each column is for a specific difference between the competing targets, $\Delta = s_A - s_B$, given in Å. The rate production of the correct product $R_A$ is a sum of a Gaussian centered at $d = 0$, which arises from the specific binding, and a uniform contribution due to non-specific binding (top row, blue). Similarly, the incorrect rate production, $R_B$, is composed of a Gaussian centered at $d = \Delta$ and a uniform non-specific contribution (top row, red). The specificity, $\xi$, is the ratio between the correct and incorrect production rates and therefore depends on the location and width of the recognition windows (bottom row). **(A)** If the competing targets differ in structure, $\Delta = 3$Å, the windows of recognition partly overlap and the resulting specificity is optimal at a *nonzero mismatch*. **(B)** For $\Delta = 0$, both $R_A$ and $R_B$ are centered around zero mismatch and the resulting specificity is approximately a rectangular window of width, $d_{1/2} \approx k^{-1/2}((N-m)\varepsilon - f - \log(\alpha r))$. **(C)** If the competing targets differ much, $\Delta \gg d_{1/2}$, the recognition windows do not overlap and the specificity is again optimal for zero mismatch. The parameters of the plot are $N = 15$, $m = 2$, $\varepsilon = 2\ k_B T$, $r = 0.1$, $g = 1$Å, $k = 1\ k_B T/\text{Å}^2$, $f = 15\ k_B T$.

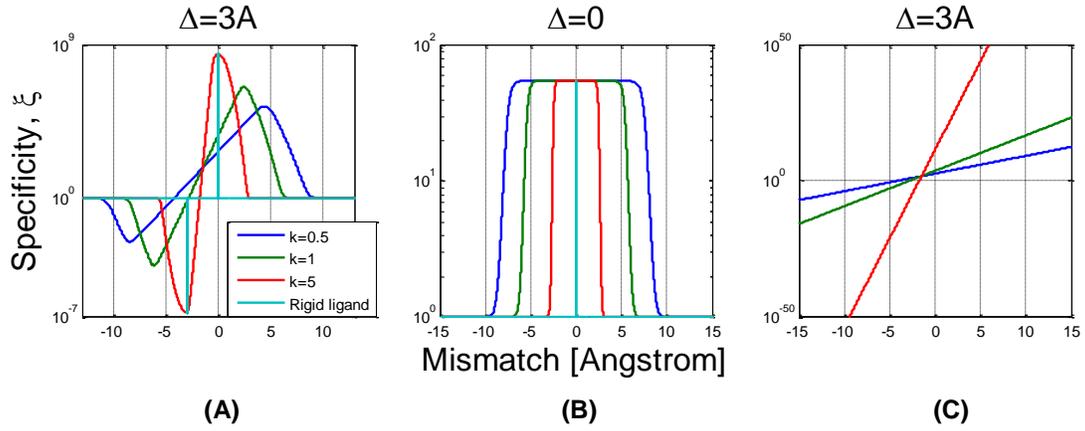

**Figure 6: Specificity ξ as a function of mismatch *d* and flexibility, *k*.** Colors denote various values of rigidity *k* (in units of $k_B T/\text{Å}^2$, legend). **(A)** For targets that differ by $\Delta = 3\text{Å}$ the specificity is optimal at a nonzero mismatch. As the ligand becomes more rigid the optimal mismatch tends to zero as $d_0 \sim k^{-1/2}$. **(B)** For competing targets with similar structure, $\Delta=0$, the specificity resembles a rectangular window centered on zero mismatch. The width of this window also decreases as $k^{-1/2}$. **(C)** The specificity when only matched complexes are functional, *r=0*, increases exponentially with the mismatch as $\xi \sim \exp(k \cdot \Delta \cdot d)$. The parameters of the plot are $N = 15$, $m = 2$, $\varepsilon = 2\, k_B T$, $r = 0.1$, $g = 1\text{Å}$, $f = 15\, k_B T$.
19

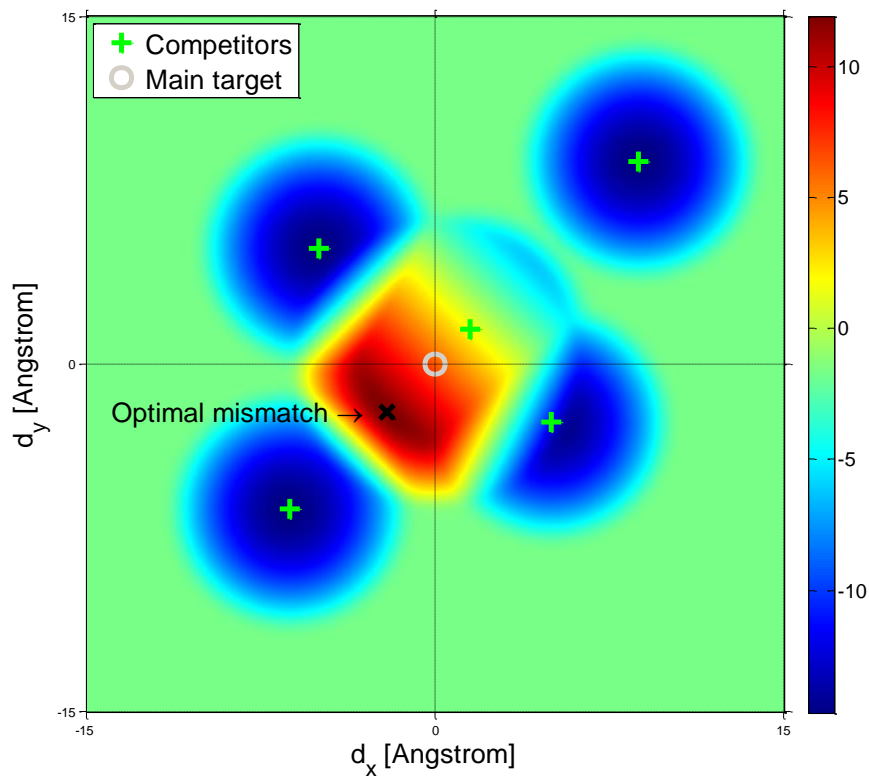

**Figure 7: Specificity as a function of a 2D mismatch in the presence of multiple competitors.** Color bar shows log of specificity. In two dimensions, the ligand structure may stretch or shrink along *x* and *y*. The mismatches along these axes are $d_x$ and $d_y$. The gray circle denotes zero mismatch. In the presence of multiple competitors (green crosses), the optimal mismatch (black X) is nonzero and depends on the structure of the various competitors. Competitors that slightly differ from the correct target have a "window of recognition" which overlaps with correct target recognition window. As a result, the specificity is maximal for a non-zero mismatch. The parameters of the plot are $N = 15$, $m = 2$, $\varepsilon = 2\ k_BT$, $r = 0.1$, $g = 1\text{Å}$, $k = 1\ k_BT/\text{Å}^2$, $f = 15\ k_BT$.

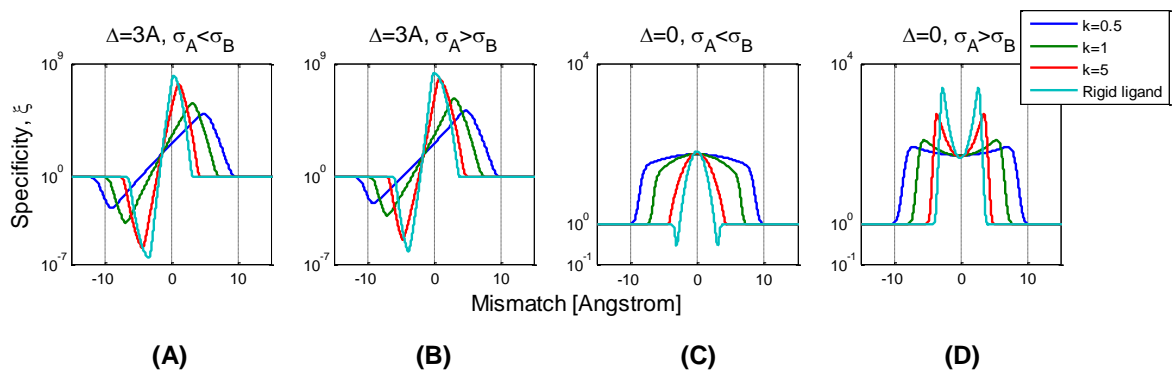

**Figure 8: Specificity ξ in the presence of two noisy targets as a function of mismatch $d$.** Colors denote various values of rigidity $k$ (in units of $k_B T/\text{Å}^2$, legend). The lengths of the target binding domain, $s_A$ and $s_B$ are fluctuating according to a Gaussian noise with variances $\sigma_A$ and $\sigma_B$. **(A, B)** For competing targets of different structure, $\Delta = 3\text{Å}$, similarly to noiseless targets, the specificity is optimal at a nonzero mismatch. **(C, D)** Competing targets of similar structure, $\Delta = 0$. The specificity has an extremum at $d = 0$, but whether this point is maximum or minimum depends on the noise. For a noisier correct target, $\sigma_A > \sigma_B$, an optimal ligand has a nonzero mismatch. The parameters of the plot are: $N = 15$, $m = 2$, $\varepsilon = 2\ k_B T$, $r = 0.1$, $g = 1\text{Å}$, $\sigma_{A,B} = 0.5, 0.6$ Å, $f = 15\ k_B T$.